\documentstyle[12pt]{article}
\include{amssymb}
\begin{document}
\newcommand{\be}{\begin{equation}}
\newcommand{\ee}{\end{equation}}
\title{Effective Levi-Civita Dilaton theory from Metric Affine Dilaton Gravity}
\author{R. Scipioni}
\maketitle
Department of Physics and Astronomy, The University of British Columbia,\\
6224 Agricultural Road, Vancouver, B.C., Canada V6T 1Z1 \footnote{scipioni@physics.ubc.ca}
\bigskip
\bigskip
\bigskip
\bigskip
\bigskip
\begin{abstract}
We show how a Metric Affine theory of Dilaton gravity can be reduced to an effective Riemannian Dilaton gravity model. A simple generalization of the Obukhov-Tucker-Wang theorem to Dilaton gravity is then presented.\\
\\
04.20.-q, 04.40.-b, 04.50.+h, 04.62.+v
\end{abstract}
\newpage
Among the four fundamental interactions, the two feeble are characterised by dimensional coupling constants, $G_{F} = (300 Gev)^{-2}$ and Newton's coupling constant $G_{N} = (10^{19} Gev)^{-2}$.\\
It is well known that interactions with dimensional coupling constants present many problems among which there is the renormalizability.\\
The success of the Weinberg-Salam model has told us that the weak interaction is characterised by a dimensionless coupling constant and the dimensions of $G_{F}$ are due to the spontaneous symmetry breaking mechanism, so that $G_{F} \cong \frac{1}{{v_{W}}^{2}}$ where $v_{W} \cong 300 Gev$ is the vacuum expectation value of the Higgs field.\\
The weakness of the weak interaction being related to the large vacuum expectation value of the scalar field [1].\\
It is believed that similar mechanisms may occur for gravity, which is characterised by a dimensionless coupling constant $\xi$. The weakness of gravity then would be related to the symmetry breaking at very high energies [2-4].\\
This is obtained starting from a Dilaton theory which presents Weyl scale invariance. The potential $V(\psi)$ which appears in the action is assumed to have its minimum at $\psi = \sigma$, then when $\psi = \sigma$ the Dilaton theory reduces to the Einstein-Hilbert action with gravitational constant $G_{N} =\frac{1}{8 \pi \xi \sigma^2}$.\\
In this Letter we investigate in the Tucker-Wang approach to non Riemannian gravity the action:
\be
S = \int k \psi^2 R \star 1 + \beta (d \psi \wedge \star d \psi) - V( \psi) \star 1
\ee
Where $R$ is the scalar curvature associated with the full non Riemannian connection.\\
In the Tucker-Wang approach to MAG we choose the metric to be orthonormal $g_{ab} = \eta_{ab} = (-1,1,1,1, ...)$ and we vary with respect to the coframe $e^{a}$ and the connection $\omega^{a}{}_{b}$ considered as independent gauge potentials.\\
As we will see the non Riemannian contribution to the Einstein-Hilbert term times $\psi^2$ is equivalent in the field equations to a kinetic term for the Dilaton and if no torsion terms are explicitly introduced in the action, the coupling $\xi$ is not arbitrary contrary to what happens in Ref [5-7] where $\xi$ is a free parameter.\\
Once $\beta$ is fixed we obtain an effective ${\xi '}^{-1}$  given by ${\xi}^{-1} + 4\frac{n-1}{n-2}$ (see eq. 22).\\ 
It has to be observed however that since we have the condition $g_{ab} = \eta_{ab}$, in general the Weyl group for the action (1) is not defined in the usual way but we may still introduce a Weyl rescaling of the form $e^{a} \rightarrow f \, e^{a}$ for the coframe, with $f$ an arbitrary function of the spacetime..\\
The variation of (1) with respect to $\psi$ gives:
\be
- \beta d \star d \psi +  k \psi (R \star 1) - V'( \psi) \star 1  = 0
\ee
By considering the variation with respect to the connection we get the equation:
\be
D \star (e_{a} \wedge e^{b}) = - \frac{2}{\psi}[d \psi \wedge \star (e_{a} \wedge e^{b})] = A(\psi)[d \psi \wedge \star (e_{a} \wedge e^{b})]
\ee
with $A(\psi) = -\frac{2}{\psi}$.\\
To prove the previous one we have to observe that the Einstein-Hilbert term which appears in the action (1) can be written as:
\be
R \star 1 = R^{a}{}_{b} \wedge \star(e_{a} \wedge e^{b})
\ee
Where $R^{a}{}_{b}$ are the curvature two forms which are defined by $R^{a}{}_{b} = d \omega^{a}{}_{b} + \omega^{a}{}_{c} \wedge \omega^{c}{}_{b}$. So we get:
\be
R \star 1 = (d \omega^{a}{}_{b} + \omega^{a}{}_{c} \wedge \omega^{c}{}_{b}) \wedge \star ( e_{a} \wedge e^{b})
\ee
Then we have to calculate the variation of:
\be
\psi ^2 \,  (d \omega^{a}{}_{b} + \omega^{a}{}_{c} \wedge \omega^{c}{}_{b}) \wedge \star ( e_{a} \wedge e^{b})
\ee
We have:
\begin{eqnarray}
\psi^2 \, (d \omega^{a}{}_{b} \wedge \star (e_{a} \wedge e^{b})) = \\ \nonumber
d [\psi^2 \, \omega^{a}{}_{b} \wedge \star (e_{a} \wedge e^{b})] +  \omega^{a}{}_{b} \wedge d[ \psi^2 \, \star (e_{a} \wedge e^{b})] \\ \nonumber
\end{eqnarray}
so (mod d):
\begin{eqnarray}
d \omega^{a}{}_{b} \wedge \star (e_{a} \wedge e^{b}) \psi^2 = \\ \nonumber
\omega^{a}{}_{b} \wedge d[\psi^2 \, \star (e_{a} \wedge e^{b})] = \\ \nonumber
\omega^{a}{}_{b} \wedge \psi^2 \, d(\star (e_{a} \wedge e^{b})) + 2 \, \psi (\omega^{a}{}_{b} \wedge d\psi \wedge \star (e_{a} \wedge e^{b})) \\ \nonumber
\end{eqnarray}
Considering the connection variation and using the definition of the covariant exterior derivative $D$ [8] we obtain formula (3).\\
The full non Riemannian Einstein Hilbert term can be written as:
\begin{eqnarray}
R \star 1 = \stackrel{o}{R} \star 1 - {\hat{\lambda}}^{a}{}_{c} \wedge {\hat{\lambda}}^{c}{}_{b} \wedge \star (e^{b} \wedge e_{a}) - \\ \nonumber 
d( {\hat{\lambda}}^{a}{}_{b} \wedge \star (e^{b} \wedge e_{a}))
\end{eqnarray}
where ${\hat{\lambda}}^{a}{}_{b}$ is the traceless part of the non Riemannian part of the connection $\lambda^{a}{}_{b}$.\\
By considering the coframe variation we get then the generalized Einstein equations:
\begin{eqnarray}
k \psi^2 {\stackrel{o}{R}}^{a}{}_{b} \wedge \star (e_{a} \wedge e^{b} \wedge e_{c}) - 2k \psi [{\hat{\lambda}}^{a}{}_{b} \wedge d \psi  \wedge \star (e^{b} \wedge e_{a} \wedge e_{c})]\\ \nonumber
- \beta [d \psi \wedge i_{c} \star d \psi  + i_{c} d \psi \wedge \star d \psi ] + k \psi^2 [{\hat{\lambda}}^{a}{}_{d} \wedge {\hat{\lambda}}^{d}{}_{b}] \wedge \star (e_{a} \wedge e^{b} \wedge e_{c}) \\ \nonumber
- V(\psi) \star e_{c} = 0
\end{eqnarray}
The Cartan equation can be written as:
\be
D \star (e^{a} \wedge e_{b}) = A(\psi)[d \psi \wedge \star (e^{a} \wedge e_{b})] = F^{a}{}_{b}
\ee
To solve the previous we need the $0-forms$ $ f^{ca}{}_{b}$ defined by $F^{a}{}_{b} = f^{ca}{}_{b} \star e_{c}$.\\
We decompose the nonmetricity and torsion as:
\begin{eqnarray}
Q_{ab} = {\hat{Q}}_{ab} + \frac{1}{n}g_{ab}Q \\ \nonumber
T_{a} = {\hat{T_{a}}} + \frac{1}{n-1}(e_{a} \wedge T) \\ \nonumber
\end{eqnarray}
where $Q_{ab} = D g_{ab}, T_{a} = d e_{a} + \omega_{a}{}^{b} \wedge e_{b}, Q = Q^{a}{}_{a}, T = i_{a}T^{a}$.\\
We have the relations:
\begin{eqnarray}
{\hat{Q_{bc}}} = \frac{1}{n} g_{bc} (f^{d}{}_{da} + f^{d}{}_{ad}) e^{a} - (f_{bac} + f_{bca} - f_{abc}) e^{a} \\ \nonumber
\hat{T_{c}} = \frac{1}{n-1} (e_{c} \wedge e^{a}) f^{d}{}_{ad} - \frac{1}{2}(e^{b} \wedge e^{a})(f_{bac} + f_{bca} + f_{cab}) \\ \nonumber
T - \frac{n-1}{2n} Q = \frac{1}{n (n-2)} (f^{c}{}_{ac} + (1-n) f^{c}{}_{ca}) e^{a}
\end{eqnarray} 
We get:
\be
f_{cab} = A(\psi) i_{c}( \star (d \psi \wedge \star (e_{a} \wedge e_{b})))
\ee
from which we get:
\begin{eqnarray}
{\hat{Q}}^{ab} = 0 \\ \nonumber
{\hat{T_{c}}} = 0
\end{eqnarray}
and
\be
T = \frac{n-1}{2n} Q + \frac{1-n}{n-2} A(\psi) d \psi
\ee
the solution for the nonmetricity and torsion can then be written as:
\begin{eqnarray}
Q_{ab} = \frac{1}{n} g_{ab} Q \\ \nonumber
T^{a} = \frac{1}{2n}(e^{a} \wedge Q) - \frac{1}{n-2}(e^{a} \wedge d \psi) A(\psi)
\end{eqnarray}
Using the expression of $\lambda^{a}{}_{b}$ as a function of $T^{a}$ and $Q_{ab}$:
\be
2 \lambda_{ab} = i_{a}T_{b} - i_{b} T_{a} -(i_{a}i_{b} T_{c} + i_{b}Q_{ac} - i_{a} Q_{bc}) e^{c} - Q_{ab}
\ee
we get:
\be
\lambda_{ab} = -\frac{1}{2n}g_{ab} Q + \frac{1}{n-2} A(\psi)(i_{a} (d \psi) e_{b} - i_{b} (d \psi) e_{a})
\ee
and the traceless part:
\be
{\hat{\lambda}}_{ab} = \frac{1}{n-2} A(\psi)(i_{a} (d \psi) e_{b} - i_{b} (d \psi) e_{a})
\ee
By using the previous expression in the generalised Einstein equations we get after some calculations:
\be
k  \psi^2 {\stackrel{o}{G}}_{c} - \beta' [d \psi \wedge i_{c} \star d \psi + i_{c} d \psi \wedge \star d \psi] -  V( \psi) \star e_{c}  = 0
\ee
where ${\stackrel{o}{G}}_{c} = {\stackrel{o}{R}}^{a}{}_{b} \wedge \star (e_{a} \wedge e^{b} \wedge e_{c})$ and:
\be
\beta' = \beta + 4k \frac{n-1}{n-2}
\ee
Then if we choose:
\be
\beta = - 4k \frac{n-1}{n-2}
\ee
we get the generalized Einstein equations reduced to:
\be
k \psi^2 {\stackrel{o}{G}}_{c} - V(\psi) \star e_{c} = 0
\ee
which are equivalent to:
\be
k {\stackrel{o}{G}}_{c} - V(\psi) \psi^{-2} \star e_{c} = 0
\ee
This are the Einstein equations we would have obtained from an Einstein theory with the potential $V(\psi) \psi^{-2}$.\\
The non Riemannian contribution to the Einstein-Hilbert term is:
\be
\Delta R \star 1 = -\frac{4}{\psi^2} \frac{n-1}{n-2}(d \psi \wedge \star d \psi)
\ee
so, the equation for $\psi$ becomes:
\be
- \beta d \star d \psi + k \psi (\stackrel{o}{R} \star 1) - k \frac{n-1}{n-2} \frac{4}{\psi}(d \psi \wedge \star d \psi) - V'(\psi) \star 1 = 0
\ee
Observe the following interesting case.\\
Suppose we start from the $\beta = 0$  in the action (1) that is:
\be
S = \int k \psi^2 R \star 1 - V(\psi) \star 1 
\ee
then we get the equations:
\begin{eqnarray}
k {\stackrel{o}{G}}_{c} \psi^2 - 4k \frac{n-1}{n-2}[d \psi \wedge i_{c} \star d \psi + i_{c} d \psi \wedge \star d \psi] - V(\psi) \star e_{c} = 0 \\ \nonumber
+ k \psi \stackrel{o}{R} \star 1 - \frac{4}{\psi} \frac{n-1}{n-2}(d \psi \wedge \star d \psi) -V'(\psi) \star 1 = 0 
\end{eqnarray}
For $n=4$ we get:
\begin{eqnarray}
k {\stackrel{o}{G}}_{c} \psi^2 - 6 k [d \psi \wedge i_{c} \star d \psi + i_{c} d \psi \wedge \star d \psi] - V(\psi) \star e_{c} = 0 \\ \nonumber
+ k \psi \stackrel{o}{R} \star 1 - \frac{6}{\psi} (d \psi \wedge \star d \psi) - V'(\psi) \star 1 = 0 
\end{eqnarray}
The Einstein equations in this case coincide formally with the conformally invariant Einstein equations obtained starting from the action:
\be
S = \int k \psi^2 \stackrel{o}{R} \star 1 + 6 k (d \psi \wedge \star d \psi) - V(\psi) \star 1
\ee
We have to remember however that this equivalence holds with the amendment that the Weyl rescaling is defined for the coframe and not for the metric since $g_{ab}$ is fixed to be orthonormal.\\
What found above can be extended to more general actions like for example:
\bigskip
\be
S = \int k \psi^2 R \star 1 + \beta (d \psi \wedge \star d \psi) + f_{1}(\psi)\frac{\alpha}{2}(dQ \wedge \star dQ) + f_{2}(\psi)\frac{\gamma}{2}(Q \wedge \star Q) - V(\psi) \star 1 
\ee
where $Q$ is the trace of the nonmetricity 1-forms $Q = g_{ab}Q^{ab}$ and $f_{1}(\psi), f_{2}(\psi)$ are two arbitrary functions of the Dilaton field $\psi$.\\
We get in this case the generalized Einstein equations:
\begin{eqnarray}
k \psi^2 {\stackrel{o}{G}}_{c} - \beta' [d \psi \wedge i_{c} \star d \psi + i_{c} d \psi \wedge \star d \psi] - V(\psi) \star e_{c} + \\ \nonumber
-f_{1}(\psi)\frac{\alpha}{2}[dQ \wedge i_{c} \star dQ + i_{c} dQ \wedge \star dQ] + f_{2}(\psi)\frac{\gamma}{2}[Q \wedge i_{c} \star Q - i_{c} Q \wedge \star Q] = 0
\end{eqnarray}
we get the equation for $Q$.\\
\be
\alpha d(f_{1}(\psi) \star d Q) + f_{2}(\psi)\gamma \star Q = 0
\ee
Eq. (33) can be considered as a generalization of the Obukhov-Tucker-Wang theorem [9-11] to the Dilaton Gravity action (32).\\
The equation for $\psi$ (27) contains other two terms which are:
\be
\frac{\alpha}{2} f_{1}'(\psi)(dQ \wedge \star dQ) + \frac{\gamma}{2}f_{2}'(\psi)(Q \wedge \star Q)
\ee
\\
If $\beta = -4k \frac{n-1}{n-2}$ then the Einstein equations would reduce to:
\begin{eqnarray}
k \psi^2 {\stackrel{o}{G}}_{c} - \\ \nonumber 
V(\psi) \star e_{c} -f_{1}(\psi)\frac{\alpha}{2}[dQ \wedge i_{c} \star dQ + i_{c} dQ \wedge \star dQ] \\ \nonumber
+ f_{2}(\psi)\frac{\gamma}{2}[Q \wedge i_{c} \star Q - i_{c} Q \wedge \star Q] = 0
\end{eqnarray}
\bigskip
The inclusion of torsion terms in the action like $T \wedge \star T$ and $T^{c} \wedge \star T_{c}$ would complicate the analysis since in that case the traceless part of the Cartan equation and then the expression for $\lambda^{a}{}_{b}$ would be modified [10], the study of these more general cases as well as the study of further possible generalizations of Obukhov-Tucker-Wang theorem will be considered in following investigations.\\
\\
\\
\\
I wish to thank the International center for cultural cooperation and development (NOOPOLIS) Italy for partial financial support, and C. Wang and R. Tucker (Lancaster) for stimulating discussions on the topic.\\
\newpage
\begin{center}
{\bf REFERENCES}
\end{center}
\bigskip
1] S. Weinberg, Phys. Rev. Lett. {\bf 19} 1264 (1967).\\
\\
2] A. Zee, Phys. Rev. Lett. {\bf 42}, 417 (1979).\\
\\
3] L. Smolin, Nucl. Phys. B {\bf 160} 223 (1979).\\
\\
4] S. L. Adler, Rev. Mod. Phys. {\bf 54}, 729 (1982).\\
\\
5] J. Kim, C. J. Park, and Y. Yonn, Phys. Rev. D {\bf 51} 562 (1995).\\
\\
6] J. Kim,, C. J. Park and Y. Yoon, Phys. Rev. D {\bf 51} 4595 (1995).\\
\\
7] Y. Yoon, gr-qc/9904018 (to appear in Phys. Rev. D).\\
\\
8] I. Benn, R. W. Tucker, \emph{An Introduction to Spinors and Geometry} (1987), Adam Hilger.\\
\\
9] R. Tucker, C. Wang, \emph{Non Riemannian Gravitational interactions}, Institute of Mathematics, Banach Center Publications, Vol. 41, Warzawa (1997).\\
\\
10] T. Dereli, M. Onder, J. Schray, R. Tucker, C. Wang, Class. Quant. Grav. {\bf 13} (1996) L103.\\
\\
11] Y. Obukhov, E. Vlachynsky, W. Esser, F. Hehl, Phys. Rev. D {\bf 56} 12, (1997) 7769.\\

\end{document}